\RequirePackage{graphicx}
\documentclass[aps,pra,floatfix,reprint,longbibliography]{revtex4-2}
\usepackage{amssymb}
\usepackage{enumitem}
\usepackage{mathtools}
\usepackage{xcolor}
\usepackage{float}
\usepackage{wrapfig}
\usepackage{graphicx}
\usepackage{natbib}
\usepackage{dcolumn}
\usepackage{bm}
\usepackage{mathtools}
\usepackage{braket}
\usepackage[utf8x]{inputenc}
\usepackage{amsmath}
\usepackage{multirow}
\usepackage{amsfonts}
\usepackage{changepage}
\usepackage{amssymb}
\usepackage{subcaption}
\usepackage{rotating} 
\usepackage{caption}
\DeclareMathOperator{\Tr}{Tr}
\usepackage{amsthm}
\theoremstyle{plain}
\theoremstyle{definition}

\newcommand{\fGam}[4]{\hat{b}^\dagger_{#1} \hat{b}^\dagger_{#2} \hat{b}^{}_{#3} \hat{b}^{}_{#4}}

\begin{document}

\title{Quantum Simulation of Bosons with the Contracted Quantum Eigensolver}



\author{Yuchen Wang, LeeAnn M. Sager-Smith, and David A. Mazziotti}
\email[]{damazz@uchicago.edu}
\affiliation{Department of Chemistry and The James Franck Institute, The University of Chicago, Chicago, IL 60637}%
\date{Submitted July 10, 2023}
%

\begin{abstract}
Quantum computers are promising tools for simulating many-body quantum systems due to their potential scaling advantage over classical computers. While significant effort has been expended on many-fermion systems, here we simulate a model entangled many-boson system with the contracted quantum eigensolver (CQE). We generalize the CQE to many-boson systems by encoding the bosonic wavefunction on qubits.  The CQE provides a compact ansatz for the bosonic wave function whose gradient is proportional to the residual of a contracted Schr{\"o}dinger equation. We apply the CQE to a bosonic system, where $N$ quantum harmonic oscillators are coupled through a pairwise quadratic repulsion.  The model is relevant to the study of coupled vibrations in molecular systems on quantum devices.  Results demonstrate the potential efficiency of the CQE in simulating bosonic processes such as molecular vibrations with good accuracy and convergence even in the presence of noise.
\end{abstract}

\maketitle

\section{Introduction}

Quantum computers have the potential to surpass classical computers in simulating quantum many-body systems~\cite{Feynman, lloyd, abrams1997,aspuru2005,lanyon2010,mcardle2020}. There has been tremendous effort to implement simulation algorithms that run on noisy intermediate-scale quantum (NISQ) devices~\cite{kitaev1995, abrams1999, dobvsivcek2007, paesani2017, peruzzo2014, wecker2015, mcclean2016, kandala2017,wang2019,tang2021,higgott2019,grimsley2019,fedorov2022}. Among them, the simulation of fermions is specifically interesting because electrons are fermions and hence, their simulation is directly related to molecular behavior. In this paper, however, we focus on many-boson systems~\cite{Huh.2015, Wang.2020, Ollitrault.2020, Lotstedt.2021, Wang.2023q63, Kovyrshin.2023, Benavides-Riveros.2020, Benavides-Riveros.2021} that are equally important in nature.

Computing the quantum energies of bosons is a fundamental problem in quantum mechanics, and it has various applications in different domains.  Applications include the study of atoms and molecules~\cite{alon2008}, condensed-matter systems such as superfluids, superconductors, and Bose-Einstein condensates~\cite{Zhang.2021w7s, Liu.2022, Hartke.2023, kasprzak2006, byrnes2014, safaei2018}, light and other optical phenomena~\cite{Mitchell.2003, Balili.2007}, chemical reactions~\cite{lin2020}, and quantum devices~\cite{Devoret.2013}.  In principle, the simulation of bosons should be more straightforward than fermions since they do not require the additional encoding that accounts for fermion antisymmetry. However, since qubits are hard-core bosons, which means they cannot occupy the same orbital, simulation of bosons is not exactly the same as simulating qubit particles. Here we employ a straightforward encoding that represents each bosonic orbital with $N$ qubits where $N$ is the number of bosons, which leads to linear scaling in both $N$ and the number of qubits.

The extension to bosonic systems in this work is achieved with the contracted quantum eigensolver (CQE)~\cite{mazziotti2021}. CQE is a general algorithm for solving the many-body Schr{\"o}dinger equation on a quantum computer. It is inspired from the contracted Schr{\"o}dinger equation (CSE), which projects the Schr{\"o}dinger equation onto a two-particle basis~\cite{nakatsuji1996, mazziotti1998, mukherjee2001, Alcoba.2001, mazziotti2006}. The key challenge in solving the CSE is to ensure the $N$-representability, that the two-electron reduced density matrix (2-RDM) must represent an $N$-particle wavefunction, which can be achieved on a classical computer with approximation techniques such as the cumulant expansion~\cite{mazziotti1999,mazziotti2000} or on quantum computer with tomographic measurements~\cite{mazziotti2021, Mazziotti.2021,Boyn.2021u94,Smart.2022l2,Smart.2023,Smart.2022,Smart.2022w8u}.  The method has been applied to calculate ground- and excited-state energies and properties with accurate results~\cite{snyder2011,boyn2021}.

In this paper, we apply the CQE to a system that consists of multiple one-dimensional harmonic oscillators. Given that the harmonic oscillator provides the fundamental model for quantum vibrations, we can consider the system where $N$ quantum harmonic oscillators are coupled by a pairwise potential as a basic approximation to coupled molecular vibration modes~\cite{sage1970coupledho, pruski1972, cohen1985,cioslowski1987,mancini1999,mazziotti1999RHI1,mazziotti2000RHI2}. The elegance of this system resides in the fact that an analytical solution can be obtained through a normal coordinate transformation and can thus be directly used to benchmark the performance of CQE on noiseless simulators as well as on NISQ devices. Using this prototype, the energies are computed with CQE and benchmarked with exact solutions.

\section{Theory}

We review the quantum algorithm for solving the anti-Hermitian part of the CSE (ACSE) in section~\ref{sec:ACSE} and introduce the bosonic mapping used in this work in section~\ref{sec:encode}. The coupled harmonic oscillator system is presented in section~\ref{sec:Hamiltonian}.

\subsection{Anti-Hermitian CSE}\label{sec:ACSE}

Consider a quantum system of $N$ identical bosons in $r$ orbitals described by the Schr{\"o}dinger equation
\begin{equation}
{\hat H} | \Psi \rangle = E | \Psi \rangle.
\end{equation}
Here $E$ and $| \Psi_{n} \rangle$ are the many-boson ground-state energy and wave function, and ${\hat H}$ is the Hamiltonian operator. One can write the Hamiltonian in second-quantized form as
\begin{equation}\label{H_sq}
{\hat H} = \sum_{pqst}{ ^{2} K^{pq}_{st} \fGam{p}{q}{t}{s}}
\end{equation}
in which $^{2} K$ is the reduced Hamiltonian matrix, the indices ranging from one to $r$ denote the orbitals. The ${\hat b}^{\dagger}_{p}$ and ${\hat b}_{p}$ are the bosonic creation and annihilation operators with respect to the $p^{\rm th}$ orbital. Taking the expectation value of the above equation yields the energy as a function of the two-particle reduced density matrix $^2 D$ (2-RDM).
\begin{equation}\label{eq:E_sq}
{E=\sum_{pqst}{^{2} K^{pq}_{st}}\langle \Psi | \fGam{p}{q}{t}{s} | \Psi \rangle}=\Tr[^2 K ^2 D]
\end{equation}
where
\begin{equation}\label{eq:D2}
{^2 D^{pq}_{st}} = \langle \Psi | \fGam{p}{q}{t}{s} | \Psi \rangle.
\end{equation}

The ACSE, which has been used to solve for energies and properties of many-particle systems~\cite{mazziotti2006, Mazziotti2007_mr,Gidofalvi2007, mazziotti2021}, has the following form
\begin{equation}\label{eq:ACSE}
\langle \Psi | [ \fGam{p}{q}{t}{s},\hat{H} ] | \Psi \rangle = 0 .
\end{equation}
The ACSE can be solved for the 2-RDM without direct computation of the many-particle wavefunction, and in previous work we have presented an algorithm to solve the ACSE on quantum devices~\cite{mazziotti2021}. The method is briefly reviewed here. Consider the variational ansatz,
\begin{equation}\label{eq:A}
|\Psi_{n+1} \rangle = e^{\epsilon \hat A_{n}}|\Psi_{n} \rangle
\end{equation}
where $\hat A_{n}$ is a two-body anti-hermitian operator
\begin{equation}\label{eq:Anhat}
{\hat A_{n}} = \sum_{pqst}{ ^{2} A^{pq}_{st} \fGam{p}{q}{t}{s} } ,
\end{equation}
that constructs the unitary transformation for updating the wavefunction.  From Eq.~(\ref{eq:E_sq}), we take the derivative of energy at the $n$-th iteration with respect to $^{2} A^{pq}_{st}$
\begin{equation}\label{eq:En}
\frac{\partial E_{n}}{\partial (^{2}A^{pq;st}_{n})} = -\epsilon \langle \Psi_{n} | [ \fGam{p}{q}{t}{s},\hat{H} ] | \Psi_{n} \rangle + O(\epsilon ^{2}) ,
\end{equation}
which shows that the negative of the residual of the ACSE yields the derivative of $E_{n}$.  Following the gradient downhill, we obtain an iterative expression for the 2-RDM
\begin{equation}\label{eq:k2}
^{2} D^{pq;st}_{n+1} = \: ^{2} D^{pq;st}_{n} + \epsilon \langle \Psi_{n} | [ \fGam{p}{q}{t}{s},\hat{H} ] | \Psi_{n} \rangle + O(\epsilon ^{2}) ,
\end{equation}
which can be expressed as a linear functional of the 1-, 2-, and 3-RDMs. On a classical computer we can approximately reconstruct the 3-RDM from the lower-order RDMs. On quantum computers we can measure the 2-RDM without explicit knowledge of the 3-RDM.  Similarly, we can measure $\hat A_{n}$ in Eq.~(\ref{eq:Anhat}) from numerical gradient techniques that are in principle exact when the step size is infinitesimal. In practical terms, the simulation and numerical gradient outcomes are collectively influenced by the step size, sampling error, and device noises. The effects of these factors will be further discussed in the results section.

\subsection{Boson to qubit mapping}\label{sec:encode}

Since qubits are hard-core bosons, we require $N$ qubits to encode the wavefunction of $N$ bosons in a single orbital. The total number of qubits required for bosonic system simulation is then equal to the number of bosons times the number of orbitals. We map the creation and annihilation operators onto qubits using the following equations, which is different from the fermionic transformation such as the Jordan-Wigner mapping~\cite{jordan_wigner_1928}:
\begin{equation}\label{eq:boson_a}
b^{\dagger}_{j,r} = \underbrace{1\otimes1\otimes...}_\text{r$\times$(j-1)+r-1} \otimes (\frac{X-iY}{2}) \otimes1\otimes...\otimes1
\end{equation}
\begin{equation}\label{eq:boson_c}
b_{j,r} = \underbrace{1\otimes1\otimes...}_\text{r$\times$(j-1)+r-1} \otimes (\frac{X+iY}{2}) \otimes1\otimes...\otimes1
\end{equation}
where $b^{\dagger}_{j,r}$ and $b_{j,r}$ represents annihilate and create the $j$-th boson in the $r$-th orbital. By summing over indices $j$, we obtain the bosonic annihilation and creation operator in $r$-th orbital
\begin{equation}\label{eq:boson_sc}
b^{\dagger}_{r} = c\sum_{j=1}^{N}b^{\dagger}_{r,j}, b_{r} = c\sum_{j=1}^{N}b_{r,j} .
\end{equation}
These operators satisfy the bosonic commutation relations when $c=1/\sqrt{N}$.

\subsection{Hamiltonian of coupled harmonic oscillators}\label{sec:Hamiltonian}

Our model system consists of $N$ spinless bosons subject to harmonic interactions in the one-dimensional space first inspected by Sage~\cite{sage1970coupledho} and later by other authors~\cite{pruski1972, cohen1985, cioslowski1987, mancini1999, mazziotti1999RHI1, mazziotti2000RHI2,Gidofalvi.2004nvn}. The one-body interactions are attractive with a scaled force constant $Z$ and the pairwise two-body interactions are repulsive. The Hamiltonian can be written as
\begin{equation}\label{H_fq}
{\hat H} = \sum_{i=1}^{N}{h(i)}+\sum_{i>j}{u(i,j)}=\sum_{i>j}{h(i,j)}
\end{equation}
where
\begin{equation}
{h(i) = -\frac{\partial^{2}}{\partial x_{i}^{2}}+Zx_{i}^{2}},
\end{equation}
\begin{equation}
{u(i,j) = - (x_{i}-x_{j})^{2}},
\end{equation}
and
\begin{equation}\label{eq:reduceH}
{h(i,j) = \frac{1}{N-1}[h(i)+h(j)]+u(i,j)} .
\end{equation}
Comparing Eqs.~(\ref{H_sq}) and~(\ref{eq:reduceH}), we obtain the matrix elements of the two-particle reduced Hamiltonian ($^2 K$)
\begin{equation}
{^{2} K^{pq}_{st}= \frac{1}{N-1}(\delta^{p}_{s}\langle q | \hat h | t\rangle + \delta^{q}_{t}\langle p | \hat h | s\rangle)+ \langle pq | \hat u | st\rangle} .
\end{equation}
\begin{figure*}[ht]
\includegraphics[scale=0.35]{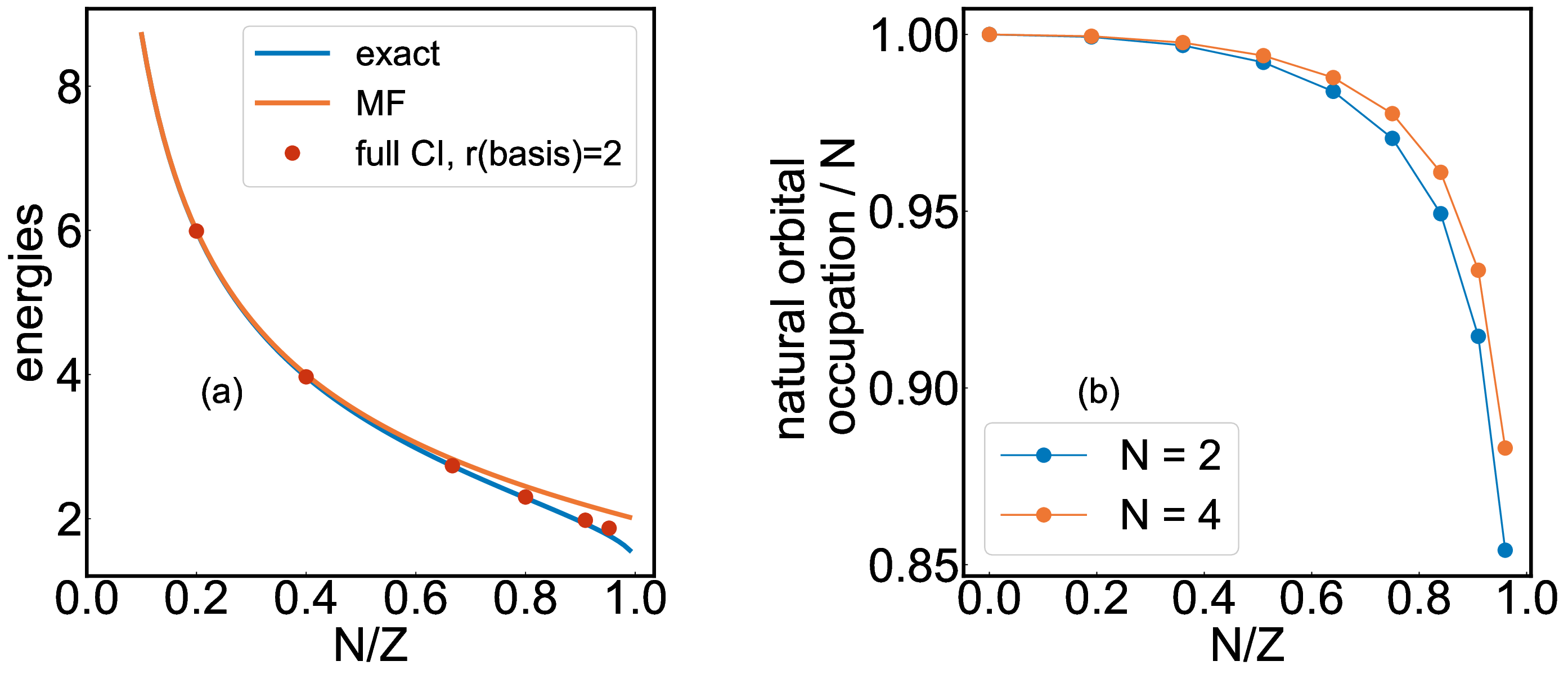}
\caption{Energies in (a) and natural orbital occupations (normalized by number of bosons) in (b) are shown as functions of $N/Z$.  MF and full CI denote mean field and full configuration interaction (exact diagonalization), and $Z$ is the harmonic force constant.} \label{fig:1}
\end{figure*}
\noindent It has been shown by Sage~\cite{sage1970coupledho} that, using the following normal coordinate transformation, we can decouple the $N$ harmonic oscillators,
\begin{equation}
Q_{i} = \begin{dcases}
      \frac{1}{\sqrt{N}} \sum_{i=1}^{N}x_{i}, \;\; i=1\\
      \frac{1}{\sqrt{i(i-1)}} \Bigr[ (i-1)x_{i}- \sum_{i=1}^{N-1}x_{i} \Bigr], \;\; 2\leq i \leq N
   \end{dcases}
\end{equation}
where the Hamiltonian in terms of these coordinates is
\begin{equation}
{\hat H} = \sum_{i=1}^{N}(-\frac{\partial^{2}}{\partial Q_{i}^{2}}) +Zx_{1}^{2}+(Z-N)\sum_{i=2}^{N}x_{i}^{2} .
\end{equation}
The system becomes $N$ uncoupled harmonic oscillators with the first one having a force constant of $\sqrt{Z}$ and the remaining $(i-1)$ indistinguishable ones having force constants of $\sqrt{Z-N}$. The system is analytically solvable and the exact ground-state energy is given as
\begin{equation}
E_{\rm exact} = \sqrt{Z} + (N-1)\sqrt{(Z-N)}
\end{equation}
which can be used to benchmark results from numerical simulations. The $n$-th natural orbital of the system has the form
\begin{equation}
\phi_{n} = (2^{n}n!\sqrt{\pi})^{-\frac{1}{2}}\gamma^{\frac{1}{2}}H_{n}(\gamma x)e^{-\gamma^2x^2/2}
\end{equation}
where $H_{n}(\gamma x)$ is the $n$-th order Hermite polynomial and $\gamma$ is a scaling factor that is determined from diagonalizing the exact 1-RDM of the system. For simulation with CQE as described below, we use the natural-orbital basis because it provides fast convergence, though other bases would also work with larger basis sizes. Note that the analytical solution in principle can be viewed as the full configuration interaction (CI) obtained from a infinite basis set. The numerical solution in a finite basis set is thus be a upper bound to the analytical solution as can be seen from Fig.~\ref{fig:1}(a).

To approach the problem from a numerical perspective, first, consider the mean-field approximation in which the energy can be easily written down as
\begin{equation}
E_{\rm MF} = N\sqrt{(Z-N+1)} .
\end{equation}
The value $N/Z$ quantifies the correlation effects. When $N/Z$ approaches zero, the two-body interaction is negligible compared to the one-body term. The system can then be viewed as $N$ almost-independent harmonic oscillators with the mean-field solution approaching the exact solution. When $N/Z$ approaches 1, the correlation effects cannot be ignored and the energy of the mean-field approximation starts to deviate from the full CI result as shown in Fig.\ref{fig:1}(a). This can also be seen from Fig.\ref{fig:1}(b), where the occupation number of the first natural orbital is plotted as a function of $N/Z$. The normalized value starts from 1, where all bosons are in the lowest orbitals, and drops significantly when $N/Z$ is close to 1, which is a result of strong correlation of the bosons. It is also worth noting that when more bosons are present, the correlation effect is reduced. This can be conceptualized as the mean-field approximation being improved by the environment that the $(N-1)$ bosons form when $N$ becomes large and eventually infinite.

For capturing the correlation effects correctly, different numerical methods have been proposed besides full CI, including the connected moments expansion (CMX) methods~\cite{cioslowski1987} and the reduced Hamiltonian interpolation (RHI)~\cite{mazziotti1999RHI1,mazziotti2000RHI2}. Discussion of these classical methods is beyond the scope of the article and is not pursued here. In this paper, we will implement CQE in strong correlated regime and benchmark with full CI and analytical exact results.

\section{Applications}

\subsection{Model Hamiltonian simulations}

We perform simulations on a quantum state-vector simulator and the IBM quantum computer Lagos. All experiments were conducted with the Qiskit package.
\begin{figure}[ht]
\includegraphics[scale=0.25]{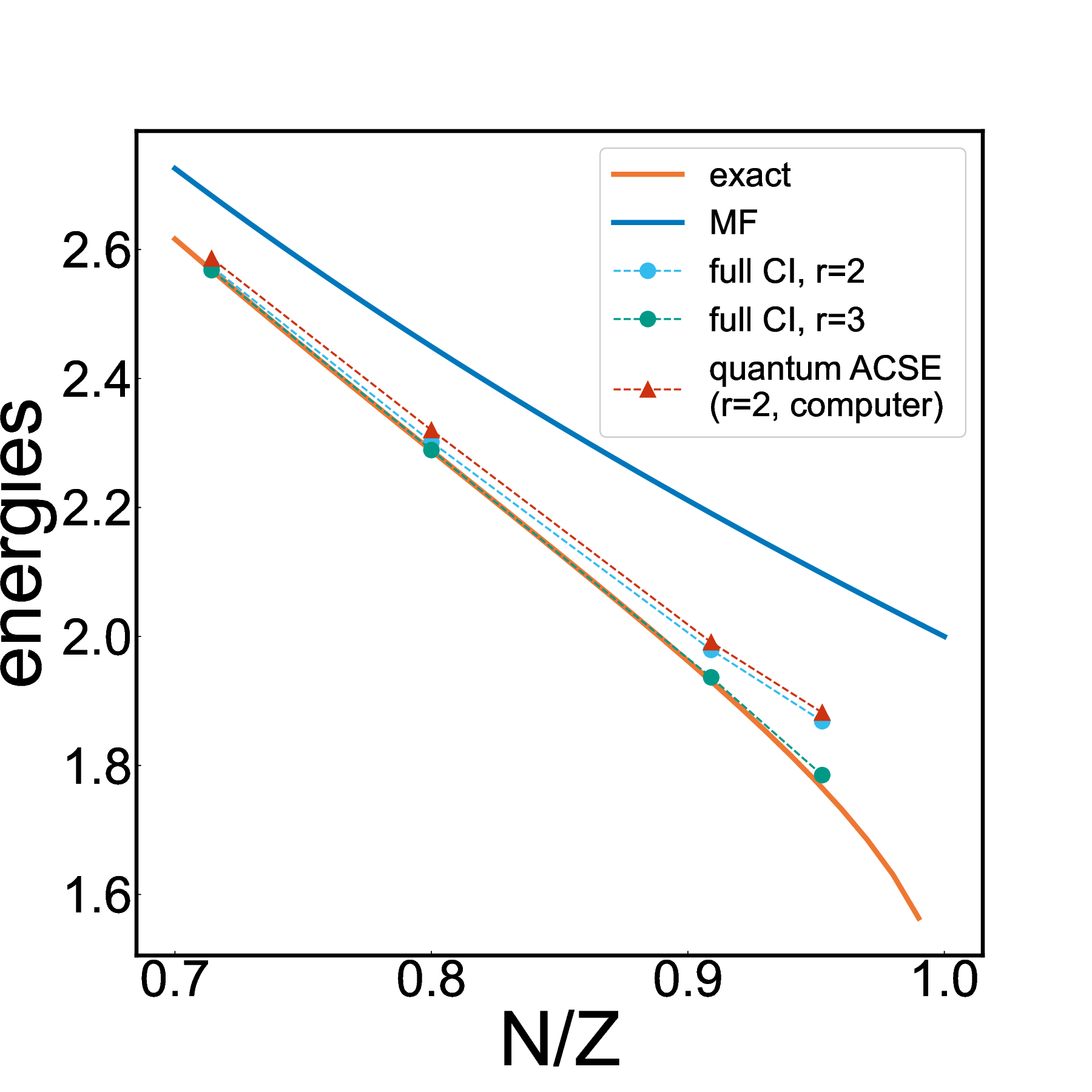}
\caption{Energies from different methods in the strongly correlated region, including the mean-field solution, the exact solution, the full CI with two-orbital basis and three-orbital bases, and a quantum ACSE (CQE) simulation performed with 2 orbitals.} \label{fig:2}
\end{figure}

We first compare the energies from different methods in the strong correlated region ($N/Z > 0.7$, $N=2$, $R=2$) in Fig.~\ref{fig:2}. The highest and lowest curves correspond to the mean-field and exact solutions, respectively. It is clear that the mean-field solution fails in the strongly correlated region. We also plot the full CI results with basis-set sizes $r=2,3$. As can be seen, CI with a three-orbital basis is sufficient to capture most of the correlation energy. The two-orbital basis performs almost as well as the three-orbital basis when $N/Z < 0.8$ but deteriorate afterwards. This basis-set truncation error exists for finite basis sets and can be improved easily with larger basis sets. The curve with triangle tickmarks shows the CQE results performed on actual quantum computers. We observe that the difference between CQE and full CI with the same basis set is much smaller than the error caused by basis-size truncation, which indicates that the accuracy of CQE is limited by the basis-set size and we expect the CQE to achieve greater accuracy with larger basis sets.

\subsection{Convergence of quantum ACSE}

In Fig.~\ref{fig:3}(a) we show a convergence diagram with two bosons in two orbitals, where a total number of four qubits are used to represent the wavefunction of the system. The starting guess places both bosons in the lowest orbital, which is the ground state in the absence of correlation. It can be seen that for both the quantum simulator and the quantum computer, the optimization converges to a stable solution in eight or fewer iterations. On an ideal state-vector simulator the error arises from the non-infinitesimal stepsize employed when evaluating $\hat A_{n}$ as well as the first-order Trotter error. Despite such error, the CQE achieves an accuracy within $~10^{-7}$ comparing to full CI, verifying the exactness of the CQE algorithm~\cite{mazziotti2021}. In the presence of device noise and sampling error, the results on a real quantum computer are only $\sim$0.01 higher than those on the noise-free simulator. This error is also fairly uniform across the iterations of the optimization.

\begin{figure}[hb]\label{fig:3}
\begin{subfigure}[i]{0.8\linewidth}
\includegraphics[width=\linewidth]{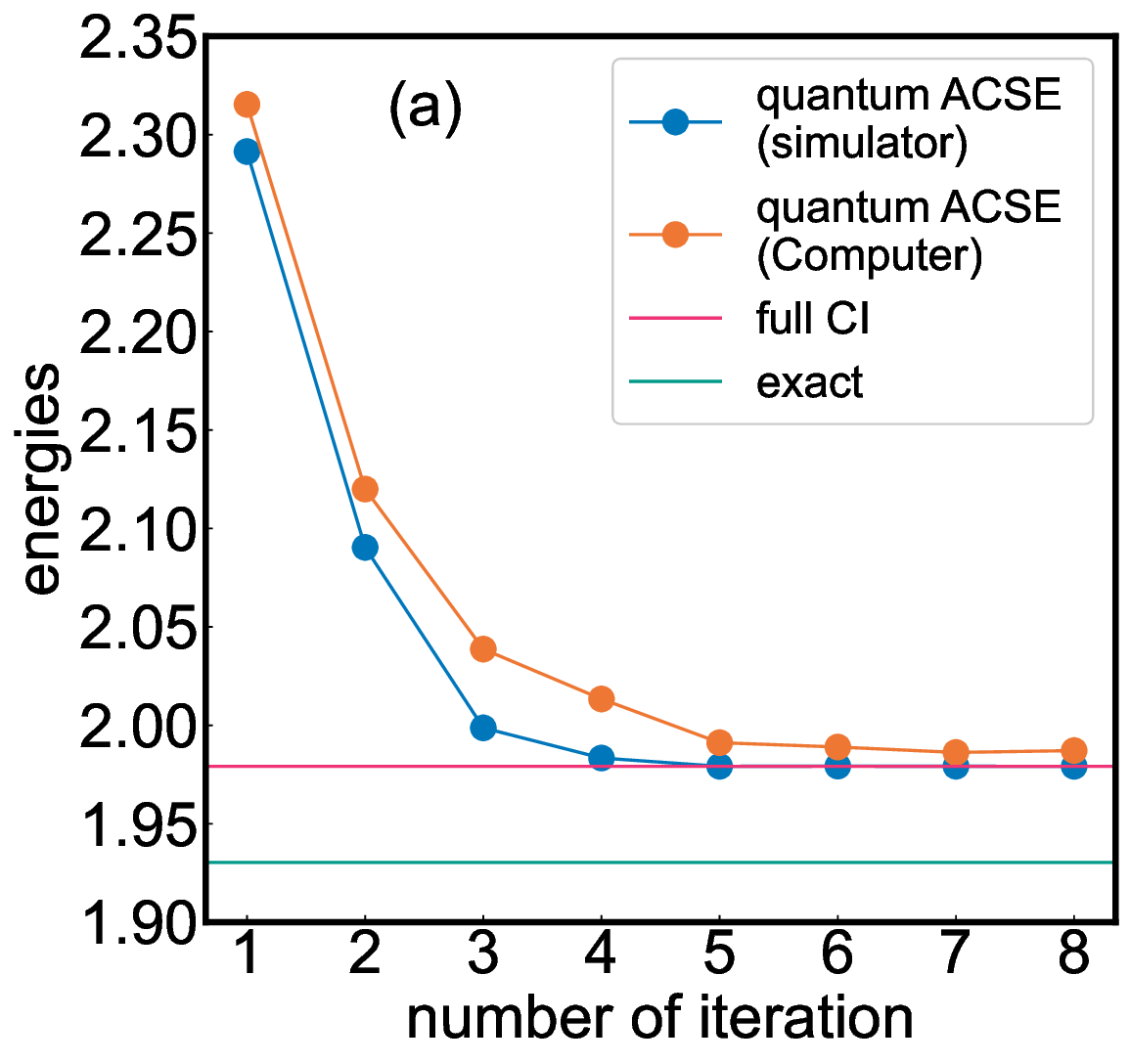}
\end{subfigure}
\hfill
\begin{subfigure}[j]{0.8\linewidth}
\includegraphics[width=\linewidth]{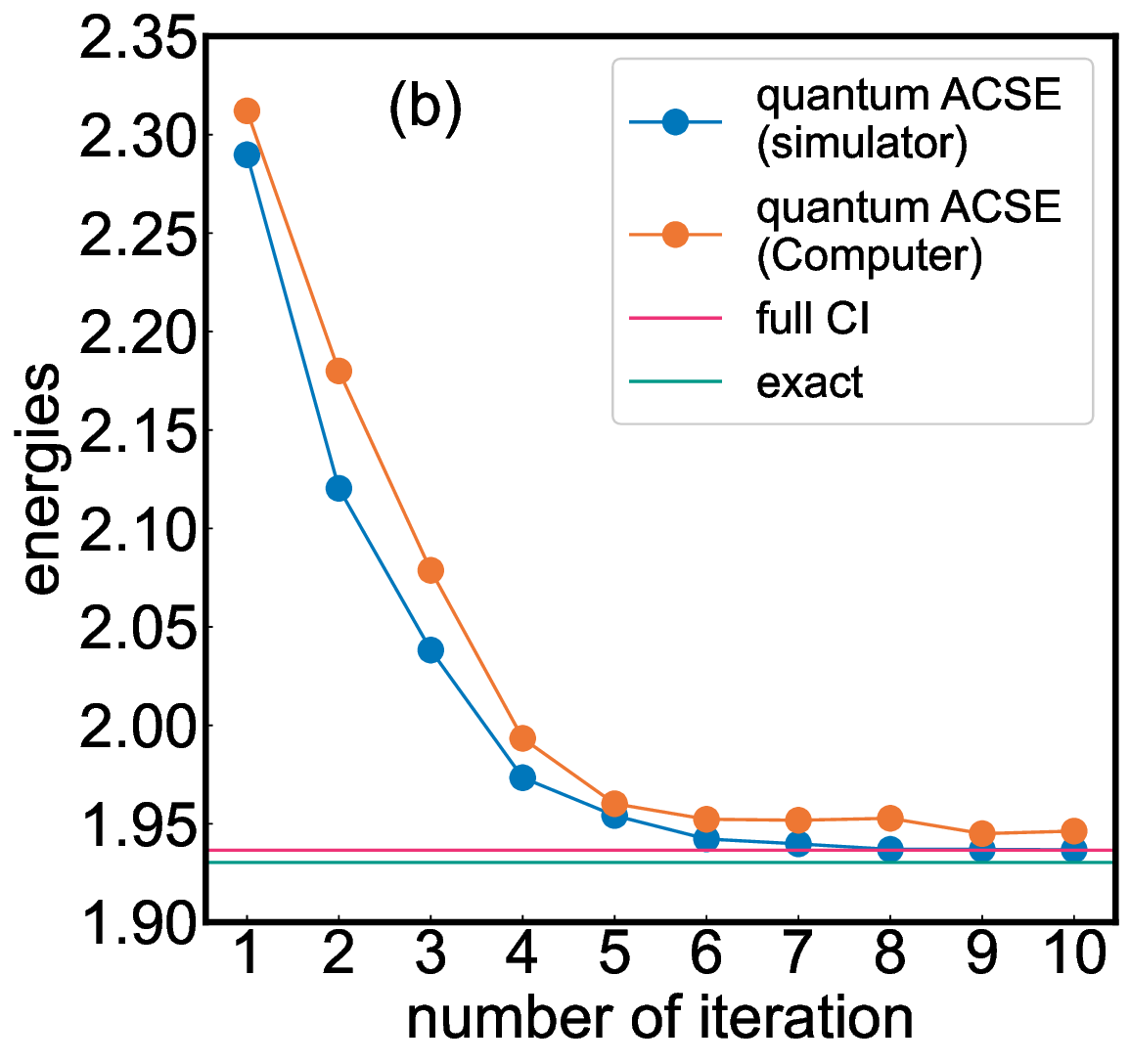}
\end{subfigure}%
\caption{Energy at each iteration when solving ACSE on simulator and real devices. Part (a) shows 2 bosons in 2 orbitals and part (b) shows 2 bosons in 3 orbitals.Z=2.2}\label{fig:3}
\end{figure}
Second, we study the effect of basis-set size by including an additional basis function. In Fig.~\ref{fig:3}(a), the CQE results are close to those from the full CI with two orbitals, but are relatively different from those from the calculation in the infinite-orbital basis set. By increasing the basis-set size by one, we observe in Fig.~\ref{fig:3}(b) that our CQE result is much closer to the exact result with a difference of ~0.015. This demonstrates that the error in Fig.~\ref{fig:3}(a) comes mostly from the finite basis set and that the CQE can reproduce a low error relative to full CI with larger basis-set sizes.

\begin{figure}[h]
\includegraphics[scale=0.35]{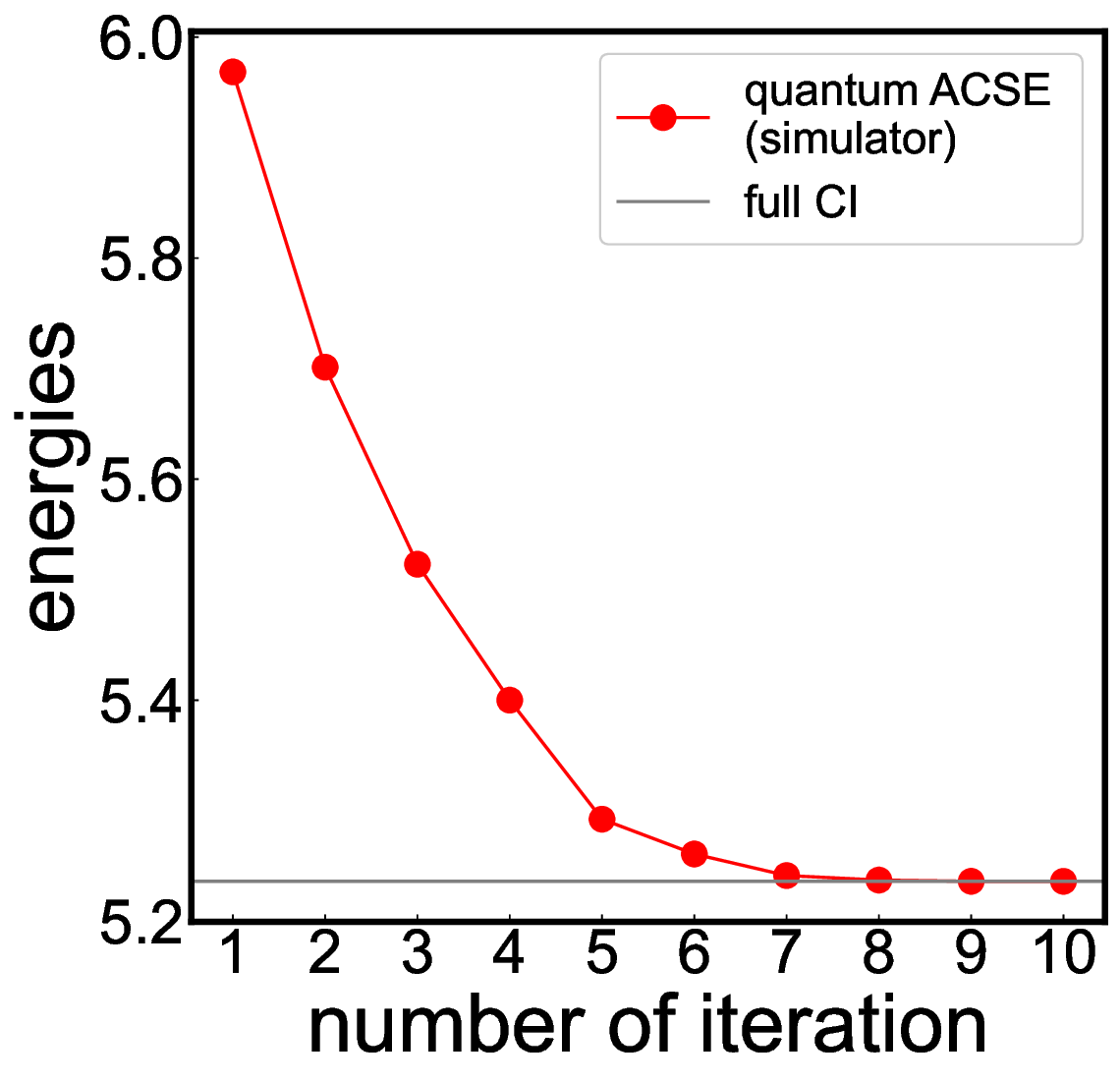}
\caption{Energy at each iteration in the solution of the ACSE (4 boson and 2 orbitals used, Z=5)} \label{fig:4b2o}
\end{figure}

Third, we perform a simulation with four bosons in two orbitals on the quantum simulator. This calculation shows that the algorithm can be straightforwardly extended to more bosons on noise-free simulators while still preserving good accuracy. The convergence diagram is plotted in Fig.~\ref{fig:4b2o}. As can be seen, the CQE still offers a near-exact solution on ideal simulators in approximately ten iterations. For molecules with more vibrational degrees of freedom, a linearly increasing number of qubits should fulfill the task, which is advantageous over certain classical algorithms.

Lastly, we explore the ability of CQE in calculating excited states. Noted that in the derivation of the ACSE in Section~IIA, we do not explicitly require the energy to be the ground-state energy. Indeed, the system of differential equations in Eqs.~\ref{eq:En} and~\ref{eq:k2} is capable of producing energy and 2-RDM solutions of the ACSE for both ground and
excited states~\cite{wang2023}. Even though excited states correspond to local stationary points rather than global energy minima of the optimization, they can be obtained from a reasonable starting guess, usually at the level of a Hartree-Fock calculation. In Table~I, we report the ground- and first excited-state energies, obtained from full CI and CQE simulation. It can be seen that performance of CQE is quite uniform in both ground- and excited-state optimization, giving an error less than 0.01 for every selected point (except for the excited state at N/Z =0.8). Here we show the CQE can be extended to treat excited states, and in future work we will show that excited states and ground state can be treated on an equal footing~\cite{yarkony2019,wang2021} on quantum devices.
\begin{table}[h]
    \caption{Full CI and CQE(computer) energy of the ground- and first-excited state for the coupled harmonic oscillator system ($N=2$,$ R=2$). For ground and excited ACSE calculation, a starting guess of $|0101\rangle$ and $|1001\rangle$ are used respectively}

    \centering
    \begin{tabular}{ccccccccc}
    \hline\hline
          N/Z & \multicolumn{2}{c}{ground} & \multicolumn{2}{c}{excited} &\\ \hline
          &  exact & CQE & exact & CQE &\\
        0.2 & 5.990719 & 5.998725 & 11.665688 & 11.672398 & \\
        0.4 & 3.968379 & 3.972838 & 7.492909 & 7.498718 & \\
        0.8 & 2.367377 & 2.370212 & 4.221183 & 4.231868 & \\ \hline
    \end{tabular}
    \label{tab:resource_count}
\end{table}

\section{Conclusions}

The simulation of bosonic systems is a fundamental problem in quantum mechanics with applications ranging from atomic and molecular physics, condensed-matter physics, and particle physics to quantum optics, quantum chemistry, and quantum computing.  We generalize the CQE to simulate many-boson systems without losing generality in treating strong correlation.  In particular, we simulate a molecular vibrational problem by solving the ACSE of the system with CQE.   The many-boson wavefunction is encoded to qubits by grouping $N$ qubits to represent a single bosonic orbital, where $N$ is the the number of bosons in the system. The total number of required qubits thus scales linearly with the number of both bosons and orbitals.

We apply CQE to a system of vibrationally coupled quantum harmonic oscillators.  Quantum harmonic oscillators---each of which is a good approximation to a single molecular vibration---are coupled by a pairwise potential to account for the coupled vibrational modes. This model can be used to treat coupled molecular modes when a normal-mode separation could be costly. We report simulation results on both a quantum simulator and a quantum device. On a noise-free simulator, the CQE results are almost exact, where the only errors arise from the Trotterized unitary transformation and the small but not infinitesimal stepsize taken in measuring the residual of ACSE. On NISQ devices, we achieve a very good accuracy in comparison with full CI results for both ground and excited states. Future work will further explore simulating excited states with the CQE.  The present work provides an important step towards the simulation of molecular vibrations on quantum devices.

\begin{acknowledgments}
D.A.M. gratefully acknowledges the Department of Energy, Office of Basic Energy Sciences, Grant DE-SC0019215 and the U.S. National Science Foundation Grants No. No. CHE-2155082, CHE-2035876, and No. DMR-2037783 and IBMQ. The views expressed are of the authors and do not reflect the official policy or position of IBM or the IBM Q team.
\end{acknowledgments}

\bibliography{
    1_intro.bib,
    2_quantum_ACSE,
    3_homodel_reference.bib,
    }

\end{document}